\documentclass[prl,twocolumn,superscriptaddress,floatfix]{revtex4}

\usepackage{graphicx}

\begin{document}

\title{Unusual behaviour of the ferroelectric polarization in PbTiO$_{3}$/SrTiO$_{3}$ superlattices}
\author{M. Dawber}
\affiliation{DPMC, University of Geneva, 24 Quai E.-Ansermet 1211
Geneva 4, Switzerland} \email{matthew.dawber@physics.unige.ch}
\author{C. Lichtensteiger}
\affiliation{DPMC, University of Geneva, 24 Quai E.-Ansermet 1211
Geneva 4, Switzerland}
\author{M. Cantoni}
\affiliation{CIME, EPFL, CH-1015 Lausanne, Switzerland}
\author{M. Veithen}
\affiliation{Institut de Physique, Universit\'{e} de Li\`{e}ge ,
Bat. B5 All\'{e}e du 6 Ao\^{u}t, 17 B- 4000 Sart Tilman, Belgium}
\author{P. Ghosez}
\affiliation{Institut de Physique, Universit\'{e} de Li\`{e}ge ,
Bat. B5 All\'{e}e du 6 Ao\^{u}t, 17 B- 4000 Sart Tilman, Belgium}
\author{K. Johnston}
\affiliation{Dept of Physics and Astronomy, Rutgers University, 136
Frelinghuysen Rd, Piscataway, NJ 08854-8019, USA}
\author{K.M. Rabe}
\affiliation{Dept of Physics and Astronomy, Rutgers University, 136
Frelinghuysen Rd, Piscataway, NJ 08854-8019, USA}
\author{J.-M. Triscone}
\affiliation{DPMC, University of Geneva, 24 Quai E.-Ansermet 1211
Geneva 4, Switzerland}

\begin{abstract}
Artificial PbTiO$_{3}$/SrTiO$_{3}$ superlattices were constructed
using off-axis RF magnetron sputtering. X-ray diffraction and
piezoelectric atomic force microscopy were used to study the
evolution of the ferroelectric polarization as the ratio of
PbTiO$_{3}$ to SrTiO$_{3}$ was changed. For PbTiO$_{3}$ layer
thicknesses larger than the 3-unit cells SrTiO$_{3}$ thickness used
in the structure, the polarization is found to be reduced as the
PbTiO$_{3}$ thickness is decreased. This observation confirms the
primary role of the depolarization field in the polarization
reduction in thin films. For the samples with ratios of PbTiO$_{3}$
to SrTiO$_{3}$ of less than one a surprising recovery of
ferroelectricity that cannot be explained by electrostatic
considerations was observed.
\end{abstract}

\maketitle

The construction of artificial ferroelectric oxide superlattices
with fine periodicity presents exciting possibilities for the
development of new materials with extraordinary properties and
furthermore is an ideal probe for understanding the fundamental
physics of ferroelectric materials.

The most studied system at present is BaTiO$_{3}$/SrTiO$_{3}$
\cite{Tabata1994,Ishibashi2000,Nakagawara2000,Shimuta2002,Neaton2003,Johnston2005,Jiang2003,Rios2003}.
Other combinations that have been studied include
KNbO$_{3}$/KTaO$_{3}$
\cite{Christen1996,Sigman2002,Sepliarsky2001,Sepliarsky2002},
PbTiO$_{3}$/SrTiO$_{3}$ \cite{Jiang1999}, PbTiO$_{3}$/BaTiO$_{3}$
\cite{LeMarrec2000},
PbTiO$_{3}$/PbZrO$_{3}$\cite{Bungaro2002,Bungaro2004a} and most
recently high quality tricolour superlattices of
SrTiO$_{3}$/BaTiO$_{3}$/CaTiO$_{3}$\cite{Lee05}. In
BaTiO$_{3}$/SrTiO$_{3}$, first principles studies \cite{Neaton2003}
suggest that both the SrTiO$_{3}$ and BaTiO$_{3}$ layers are
polarized such that the polarization is approximately uniform
throughout the superlattice. The driving force behind this is the
large electrostatic energy penalty for a build-up of charge at the
interface caused by discontinuous polarization in the normal
direction. The electrostatic model proposed by Neaton and Rabe
\cite{Neaton2003} to explain their first principles results for
BaTiO$_{3}$/SrTiO$_{3}$ superlattices is very similar to the
electrostatic model applied to calculate the effect of the
depolarization field in ultra-thin ferroelectric films with
realistic electrodes \cite{Batra1972,Junquera2003,Dawber2003}.
Experimentally it was recently shown that the reduced polarization
observed in monodomain thin PbTiO$_{3}$ can be explained by the
presence of a depolarization field resulting from imperfect
screening of the polarization \cite{Lichtensteiger}. Recent work
also suggests that, under certain conditions, the electrostatic
energy due to depolarization fields will drive the system to form
domains as observed by Fong et al. \cite{Fong2004} and Nagarajan et
al. \cite{Nagarajan}. In this letter we use PbTiO$_{3}$/SrTiO$_{3}$
superlattices to probe the effect of a reduced ferroelectric
thickness in a dielectric environment. Our data show that the
behaviour observed in PbTiO$_{3}$ thin films is reproduced for
PbTiO$_{3}$ layers thicker than three unit cells. However, for
thinner ferroelectric layers a surprising recovery of
ferroelectricity that cannot be explained by electrostatic
considerations was observed.

The superlattices of PbTiO$_{3}$/SrTiO$_{3}$ were prepared on
conducting 0.5$\%$ Nb doped (001) SrTiO$_{3}$ substrates using
off-axis RF magnetron sputtering with  conditions similar to those
used for growing high quality epitaxial c-axis PbTiO$_{3}$ thin
films \cite{Lichtensteiger}. For all the samples discussed in this
paper, the SrTiO$_{3}$ thickness was fixed at three unit cells
(about 12 \AA). At room temperature the in-plane lattice parameters
of tetragonal ferroelectric PbTiO$_{3}$ (a=3.904 \AA,c=4.152 \AA)
and cubic dielectric SrTiO$_{3}$ (3.905 \AA) are an excellent
match\cite{Landolt-Bornstein}. It is thus expected that PbTiO$_{3}$
will grow coherently on SrTiO$_{3}$ substrates, and that the strain
interactions will be dominated by the constraint imposed by the
substrate. The growth temperature for the superlattices was
460$^{o}$ C. Investigation by Transmission Electron Microscopy (TEM)
revealed excellent quality in superlattices with layers of
SrTiO$_{3}$ thinner than 5 unit cells, though beyond this thickness
the quality of the SrTiO$_{3}$ layers deteriorated with thickness,
presumably because of the low temperature. On the other hand,
samples processed with higher temperatures were of lower quality,
probably because of lead losses from the PbTiO$_{3}$. The low
temperature growth used thus seemed to be optimal for making
superlattices in which the SrTiO$_{3}$ layers are very thin, but
without limitation on the thickness of PbTiO$_{3}$.

In the principal series of interest we grew superlattices consisting
of 20 PbTiO$_{3}$/SrTiO$_{3}$ bilayers in which the SrTiO$_{3}$
layer thickness was maintained at 3 unit cells while the PbTiO$_{3}$
layer thickness n was varied from 54 unit cells down to just one
unit cell (denoted n/3). The first layer deposited was PbTiO$_{3}$.
The layer thicknesses were determined by calibrating x-ray
diffraction patterns with deposition time.

\begin{figure}[ht]
  \includegraphics[width=8cm]{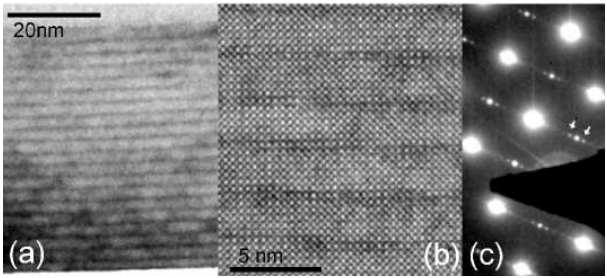}
  \caption{\textit{Cross sectional TEM images of a 20 bilayer PbTiO$_{3}$/SrTiO$_{3}$ 3/3 sample. (a) Bright field image clearly shows the intended layering of the structure. (b) HRTEM shows the perfect crystalline structure of the material. (c) Diffraction image demonstrating superlattice periodicity.}}\label{fig:TEM}
\end{figure}

Cross-sectional TEM investigations were performed on several samples
and reveal the coherent growth and artificial layering of the
samples. Fig. \ref{fig:TEM} shows a summary of the results obtained
on a 3/3 sample. The bright field image, Fig.\ref{fig:TEM} (a),
shows the layering throughout the sample. The perfect crystalline
structure and coherent growth are demonstrated by the high
resolution TEM (HRTEM) image Fig.\ref{fig:TEM} (b), while the
periodicity of the superlattice is demonstrated by the superlattice
reflections in the diffraction image Fig.\ref{fig:TEM} (c) (arrows).

\begin{figure}[ht]
  \includegraphics[width=8.5cm]{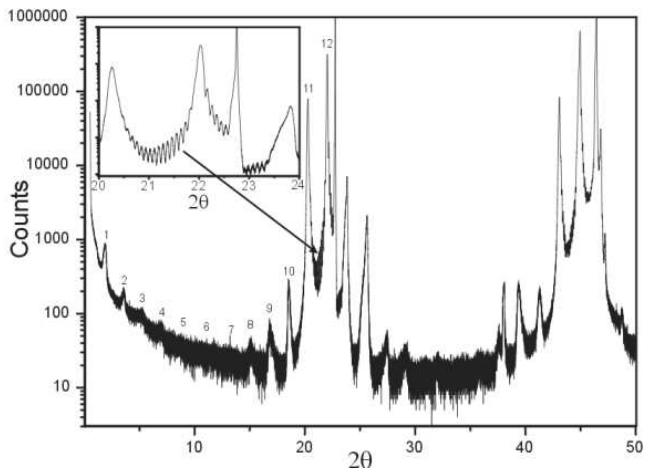}
  \caption{\textit{$\theta-2\theta$ x-ray diffractogram for a 20 bilayer PbTiO$_{3}$/SrTiO$_{3}$ 9/3 superlattice.}}\label{fig:xrays}
\end{figure}

Further structural characterization were performed using standard
$\theta-2 \theta$ x-ray diffraction. Fig. \ref{fig:xrays} shows the
$\theta-2\theta$ scan for a superlattice in which the PbTiO$_{3}$
layers are 9 unit cells thick and the SrTiO$_{3}$ layers are 3 unit
cells thick (9/3). The periodicity of the superlattice is therefore
12 perovskite unit cells and 12 reflections from $2\theta=0$ to the
angle corresponding to the 001 peak of the average perovskite unit
cell lattice parameter (at $2\theta\approx22^o$) are expected, most
of which are observable in the scan. In between the main
superlattice peaks, the presence of 18 finite size effect peaks,
clearly visible in the inset of Fig. \ref{fig:xrays}, is due to the
finite total thickness of the sample which is 20 times the
superlattice periodicity.

\begin{figure}[ht]
  \includegraphics[width=7.5cm]{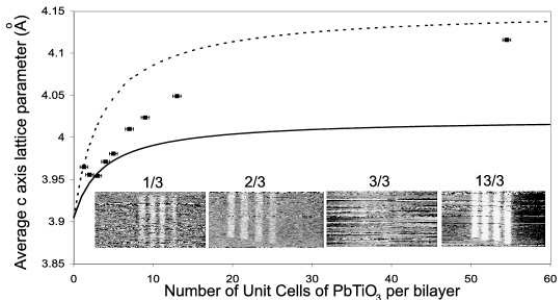}
  \caption{\textit{Average c-axis lattice parameter plotted against the number of unit cells of lead titanate per bilayer showing the suppression and recovery of ferroelectricity. Complementary PFM images are shown as insets.}}\label{fig:mainfig}
\end{figure}

Because of the large strain-polarization coupling in PbTiO$_{3}$
\cite{Cohen92}, a change in polarization results in a change in
material tetragonality \cite{Lichtensteiger}. We take advantage of
this to follow the evolution of the polarization in the superlattice
by following the evolution of the average c axis lattice parameter,
$\bar{c}$, as the PbTiO$_{3}$ layer thickness is varied. If the
wavelength of the superlattice is $n\bar{c}$ then the n$^{th}$ peak
in a $\theta-2\theta$ scan will always correspond to $\bar{c}$
irrespective of the value of $n$ allowing the average c axis lattice
parameter of the superlattice to be easily determined. In practice,
this peak is easily identifiable due to its high intensity and
proximity to the substrate peak. Intuitively one expects, as the
thickness of the PbTiO$_{3}$ layers relative to the SrTiO$_{3}$
layers is reduced, a decrease of the ferroelectric polarization
which should result in a concomitant decrease of the average lattice
parameter. The measured average c axis lattice parameters as a
function of the thickness of  the PbTiO$_{3}$ layer thickness are
shown in Fig. \ref{fig:mainfig}. For comparison, we also show the
average c axis lattice parameters obtained by fixing c of SrTiO$_3$
at its paraelectric cubic value 3.905 {\AA}  and taking c of
PbTiO$_3$ in two limiting cases: first, at the value 4.022 {\AA}
corresponding to a hypothetical paraelectric tetragonal structure
coherent with the substrate (solid line)\cite{Lichtensteiger} and
then at the fully polarized bulk value 4.152 {\AA} (dashed line). As
can be seen in Fig. \ref{fig:mainfig}, superlattices with thick
PbTiO$_{3}$ layers have ``large'' average lattice parameters clearly
suggesting a ferroelectric polarization. On reduction of the layer
thickness the average lattice parameter decreases and approaches the
solid line. However, surprisingly, after reaching this line
superlattices with very small PbTiO$_{3}$ layer thicknesses display
larger average lattice parameters which indicate a recovery of
ferroelectricity.

This behavior was confirmed using atomic force microscopy (AFM)
which allows the ferroelectric domain structure to be modified and
detected on a local scale\cite{Paruch01}. Applying a voltage between
the metallic tip of the AFM and the metallic substrate, stripes were
``written'' (poled) using alternatively positive and negative
voltages. Piezoelectric atomic force microscopy (PFM) was then used
to detect the domain structure. PFM images are shown in the insets
of Fig. \ref{fig:mainfig} for different superlattices, the contrast
revealing domains with up and down polarization. As can be seen, the
1/3, 2/3 and 13/3 samples reveal a clear domain structure and are
indeed ferroelectric whereas no significant contrast could be
obtained in the 3/3 superlattice, confirming the behavior suggested
by the x-ray analysis.  The written domains for all samples in which
domains could be written were confirmed to be stable for a number of
days. All domains written into the down direction have the same
piezoelectric response as the existing background, implying that
before writing the entire sample is uniformly poled in the down
direction, demonstrating that none of the samples formed a
polydomain state. This is a particularly important observation for
the 1/3 and 2/3 samples since a multidomain configuration could
possibly explain the observed increase in average c axis lattice
parameter at very small PbTiO$_{3}$ layer thicknesses.

To understand the observed behaviour, a simple electrostatic model
similar to the one proposed by Junquera and Ghosez
\cite{Junquera2003} has been developed. The total energy per unit
cell area $E$ of an $n_p/n_s$ superlattice is written as
\begin{equation}
\label{energy} E (P_p^0,P_s^0)= n_p \, \, \, U_p(P_p^0) + n_s \,
\, \, U_s(P_s^0) + E_{elec}(P_p^0,P_s^0),
\end{equation}
where $U_p$ and $U_s$ are the total energies per 5-atom unit cell of
bulk PbTiO$_3$ and SrTiO$_3$ in zero field as a function of their
polarization $P_p^0$ and $P_s^0$ (assumed to be homogeneous in each
layer) and $E_{elec}$ is the macroscopic electrostatic energy
resulting from the presence of non-vanishing electric fields in the
layers when $P_p^0$ and $P_s^0$ differ.

The electrostatic energy of a given layer, of thickness $l_p$ or
$l_s$, in the presence of a finite electric field $\cal{E}$, to
leading order in the field, is $ E_{elec} = - l {\cal E} \cdot
P^{0}$. In the superlattice, the electric fields ${\cal E}_p$ and
${\cal E}_s$ are determined by $P_p^0$ and $P_s^0$ through the
condition of continuity of the normal component of the electric
displacement field at the interfaces:
\begin{equation}
P_p^{0} + \varepsilon_{0} {\cal E}_p =
P_s^{0} + \varepsilon_{0} {\cal E}_s
\end{equation}
For a system under short-circuit boundary conditions, the potential
drop along the structure must vanish so that
\begin{equation}
l_p {\cal  E}_p   = - l_s {\cal  E}_s
\end{equation}
Combining the last two conditions and summing the electrostatic
energies of the PbTiO$_3$ and SrTiO$_3$ layers we obtain
 \begin{equation}
E_{elec}(P_p^0,P_s^0) = \frac{l_p l_s }{ \varepsilon_{0} (l_p + l_s) } (P_s^{0} - P_p^{0})^{2}
\end{equation}

The total energies $U_p(P_p^0)$ and $U_s(P_s^0)$ have been obtained
from density functional theory (DFT) calculations on bulk compounds
using the {\sc abinit} package \cite{abinit}. The calculations were
performed within the local density approximation (LDA) using
extended norm conserving pseudopotentials~\cite{Teter} with the Pb
(5d,6s,6p), Sr (4s,4p,5s), Ti (3s,3p,3d,4s) and O (2s,2p) treated as
valence states. Convergence was reached for a 1225 eV (45 Ha) cutoff
and a $6 \times 6 \times 6$ mesh of special $k$-points. We obtain
computed lattice constants for cubic paraelectric SrTiO$_3$
($a=3.846$ {\AA}) and for tetragonal ferroelectric PbTiO$_3$
($a=3.864$ {\AA}, $c=3.975$ {\AA}), with a polarization of 0.69
C/m$^2$. Both materials lattice parameters are underestimated
relative to the experimental values as is typical for the LDA. For
each compound, $U(P^0)$ and $c(P^0)$ were obtained~\cite{expansion}
following the formalism of Ref. \onlinecite{NaSai} by relaxing the
atomic positions and the lattice parameter $c$ at fixed polarization
$P^0\hat z$ in the space group $P4mm$, constraining the in-plane
lattice parameter $a$ to 3.846 {\AA}. For bulk PbTiO$_{3}$
constrained in plane to 3.846 {\AA} the c axis lattice parameter was
found to be 4.009 {\AA} with a polarization of 0.73 C/m$^{2}$. For
any choice of $n_p$ and $n_s$, minimization of Eq. \ref{energy}
gives equilibrium values for $P_p^0$ and $P_s^0$, and thus also for
$c_p$ and $c_s$.

To see whether the model correctly describes the behaviour as the
thickness of the PbTiO$_3$ layers decreases to the atomic scale, we
performed full DFT-LDA calculations of the structure and
polarization of PbTiO$_{3}$/SrTiO$_{3}$ superlattices for $n_s$ = 3
and $n_p$ = 1,... 7 with the Vienna ab initio Simulations Package
({\sc vasp}) \cite{Kresse1996a}, using projector augmented wave
(PAW) potentials \cite{Blochl1994a,Kresse1999a} with the same
valence configurations as in the {\sc abinit} calculation.
Convergence was reached for a 600 eV (22 Ha) cutoff and a $6 \times
6 \times 2$ mesh of special $k$-points. The computed lattice
constants are for SrTiO$_3$, $a=3.86$ {\AA}, and for tetragonal
ferroelectric PbTiO$_3$, $a=3.86$ {\AA} and $c=4.047$ {\AA}, with a
polarization of 0.75 C/m$^{2}$ \cite{calcnote}. For the
superlattices, the atomic positions and lattice parameter $c$ were
fully relaxed in the space group $P4mm$, constraining the in-plane
lattice parameter $a$ to 3.86 {\AA}. Polarizations were calculated
using the modern theory of polarization \cite{Vanderbilt1993} as
implemented in {\sc vasp}.

\begin{figure}[ht]
  \includegraphics[width=7.5cm]{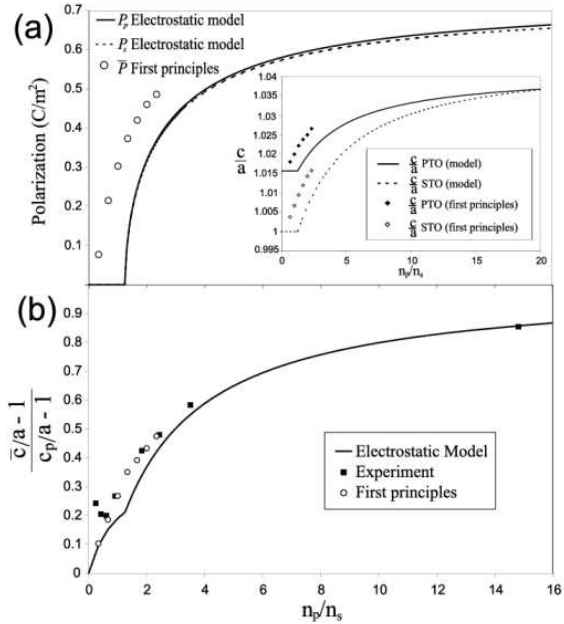}
  \caption{\textit{(a) Polarization in each layer
obtained from the electrostatic model (dotted and solid lines) and
the average polarization from first principles calculations (full
circles). Inset shows tetragonality in each material calculated from
both methods. (b) Comparison of experiment and both theoretical
approaches.}}\label{fig:theoryfigure}
\end{figure}

Fig. \ref{fig:theoryfigure} (a) shows the evolution of the
polarization as a function of $\frac{n_p}{n_s}$ for $n_s$ = 3.
According both to the model and to the first principles local
polarizations (not shown), the difference between the polarizations
in the two layers is quite small, highlighting the large
electrostatic energy cost of having different polarizations in the
layers. As the ratio $n_p/n_s$ increases, the polarization of the
superlattice asymptotically approaches the constrained bulk
PbTiO$_3$ value, though rather slowly due to the large energy cost
of maintaining a high polarization in SrTiO$_{3}$ . The figure inset
shows the corresponding increase in the tetragonality (c/a) of the
two layers, with the high polarization-strain coupling in the
SrTiO$_{3}$ layer (higher even than for the PbTiO$_3$
layer~\cite{expansion}) being evident. Both the model and the first
principles calculations show a monotonic decrease of the
polarization as the PbTiO$_{3}$ volume fraction is reduced, due to
the increase in the relative energy cost of the polarization in the
SrTiO$_{3}$ layers. While the polarization vs thickness curve for
the model at the lowest thicknesses is shifted to lower
polarizations relative to the first principles results, the model
works overall very well, considering the simplifying assumptions and
lack of any adjustable parameters.

In Fig. \ref{fig:theoryfigure} (b) we compare results from the first
principles calculations (open circles), the electrostatic model
(solid line) and experiment(solid squares), by plotting the
fractional change in the superlattice tetragonality
$\frac{\bar{c}}{a}-1$ relative to the tetragonality of bulk
PbTiO$_{3}$ with the in plane lattice parameter constrained to the
SrTiO$_{3}$ substrate \cite{calcnote}. Good agreement between both
theoretical approaches and experiment is seen for samples that are
predominantly PbTiO$_{3}$. It should be noted that both theoretical
calculations are at zero temperature, whilst the experiments are
conducted at room temperature. Specifically this means that samples
predicted from first principles to be ferroelectric with a small
polarization at zero temperature might be expected to be
paraelectric in our room temperature experiment, as is observed in
the case of the 3/3 sample. The fact that unexpected recovery of the
ferroelectric polarization in the experimental 1/3 and 2/3
superlattices is observed in neither the electrostatic model, nor
the first principles calculations, suggests that it is related to
aspects not accounted for in our theoretical approaches, for
example, the precise nature of the substrate-superlattice interface,
some degree of intermixing at the superlattice interfaces, or the
possible formation of a new entropically stabilized PbTiO$_{3}$
phase similar to that formed under negative hydrostatic pressure in
the first principles studies of Tinte et al \cite{Tinte2003}.

This work was supported by the Swiss National Science Foundation
through the National Center of Competence in Research ``Materials
with Novel Electronic Properties-MaNEP'', the VolkswagenStiftung
(project ``Nano-sized ferroelectric hybrids'', I80 899),
FNRS-Belgium (grant 2.4562.03), the European NoE FAME and ESF
(THIOX) and DOE Grant DE-FG02-01ER45937.


\begin{thebibliography}{99}

\bibitem{Tabata1994}
H.Tabata, H.Tanaka, and T.Kawai, {\sl Appl. Phys. Lett.}
  \textbf{65} 1970 (1994).

\bibitem{Ishibashi2000}
Y. Ishibashi,  N.Ohashi, and T.Tsurumi, {\sl Jap. J. Appl. Phys.}
\textbf{39} 186 (2000).

\bibitem{Nakagawara2000}
O.Nakagawara, T.Shimuta and T. Makino, {\sl Appl. Phys. Lett.}
  \textbf{77} 3257 (2000).

\bibitem{Shimuta2002}
T. Shimuta, O.Nakagawara and T.Makino, {\sl Jap. J. Appl. Phys.}
\textbf{91} 2290 (2002).

\bibitem{Neaton2003}
J.B. Neaton and K.M. Rabe, {\sl Appl. Phys. Lett.} \textbf{82} 1586
(2003).

\bibitem{Johnston2005}
K. Johnston, X. Huang, J.B. Neaton and K.M. Rabe {\sl Phys. Rev. B}
\textbf{71} 100103(R) (2005).

\bibitem{Jiang2003}
A.Q. Jiang, J.F. Scott, H. Lu and Z. Chen, {\sl J. Appl. Phys.}
\textbf{93} 1180 (2003).

\bibitem{Rios2003}
S. Rios et al., {\sl J. Phys. Cond. Matt.} \textbf{15} 305 (2003).

\bibitem{Christen1996}
H.M. Christen et al., {\sl Appl. Phys. Lett.} \textbf{68} 1488
(1996).

\bibitem{Sigman2002}
J. Sigman et al., {\sl Phys. Rev. Lett.} \textbf{88} 097601 (2002).

\bibitem{Sepliarsky2001}
M. Sepliarsky, S.R. Phillpot and D. Wolf,  {\sl J. Appl. Phys.}
\textbf{90} 4509 (2001).

\bibitem{Sepliarsky2002}
M. Sepliarsky, S.R. Phillpot, M.G. Stachiotti and R.L. Migoni, {\sl
J. Appl. Phys.} \textbf{91} 3165 (2002).

\bibitem{Jiang1999}
J.C. Jiang et al., {\sl Appl. Phys. Lett.} \textbf{74} 2851 (1999).

\bibitem{LeMarrec2000}
F. Le Marrec et al. {\sl Phys Rev B} \textbf{61} R6447 (2000).

\bibitem{Bungaro2002}
C. Bungaro and  K.M. Rabe,  {\sl Phys. Rev. B} \textbf{65} 224106
(2002).

\bibitem{Bungaro2004a}
C. Bungaro  and K.M. Rabe, {\sl Phys. Rev. B} \textbf{69} 184101
(2004).

\bibitem{Lee05}
H.N. Lee et al., {\sl Nature} \textbf{433} 395 (2005).

\bibitem{Batra1972}
I.P. Batra and B.D. Silverman, {\sl Solid State Commun.} {\bf 11}
291 (1972).

\bibitem{Junquera2003}
J. Junquera and P. Ghosez, {\sl Nature} \textbf{422} 506 (2003).

\bibitem{Dawber2003}
M. Dawber, P. Chandra, P.B. Littlewood and J.F. Scott, {\sl J. Phys.
C.} \textbf{15} L393 (2003).

\bibitem{Lichtensteiger}
C. Lichtensteiger, J.-M. Triscone, J. Junquera and P. Ghosez, {\sl
Phys. Rev. Lett.} \textbf{94} 047603 (2005).

\bibitem{Cohen92}
R. Cohen, {\sl Nature(London)} \textbf{358} 136 (1992).

\bibitem{Fong2004}
D.D. Fong et al., {\sl Science} \textbf{304} 1650 (2004).

\bibitem{Nagarajan}
V. Nagarajan, J. Junquera et al., {\sl unpublished}.

\bibitem{Landolt-Bornstein}
Landolt-B\"{o}rnstetin, III, 16a, eds. K.H. Hellwege and A.M.
Hellwege, Springer-Verlag (1981).

\bibitem{Paruch01}
P. Paruch, T. Tybell and J.-M. Triscone, {\sl Appl. Phys. Lett.}
\textbf{79} 530 (2001).

\bibitem{abinit}
X. Gonze et al., {\sl Computational Materials Science} \textbf{25},
478 (2002).

\bibitem{NaSai}
Na Sai, K. M. Rabe and D. Vanderbilt, {\sl Phys. Rev. B}
\textbf{66}, 104108 (2002).

\bibitem{Teter}
M. Teter, {\sl Phys. Rev. B} \textbf{48}, 5031(1993).

\bibitem{expansion}
The calculations provide $U(P_0) = B P_0^{2} +C P_0^{4}$ and $c/a =
\alpha + \beta P_0^{2} + \gamma P_0^{4}$. For $U$ in [eV/cell] and
$P_0$ in [C/m$^2$] : $ B_s = 0.21046331 , B_p = - 0.17175279 , C_s
=0.30913420 , C_p = +0.16068441 , \alpha_s = 1.0 , \alpha_p =
1.01566146 , \beta_s = 0.06076952, \beta_p = 0.03609915, \gamma_s =
0.04820368 , \gamma_p = 0.02209009$.


\bibitem{Blochl1994a}
P.E. Bl\"{o}chl, {\sl Phys. Rev. B} \textbf{50} 17953 (1994).

\bibitem{Kresse1996a}
G. Kresse and J. Furthm\"{u}ller, {\sl Phys. Rev. B} \textbf{54}
11169 (1996).

\bibitem{Kresse1999a}
G. Kresse and D. Joubert, {\sl Phys. Rev. B} \textbf{59} 1758
(1999).

\bibitem{calcnote}
The {\sc abinit} \cite{abinit} and {\sc vasp} calculations have been
done with different pseudopotentials, cutoffs and
exchange-correlation functionals, which collectively accounts for
the difference in the c axis lattice parameter obtained for
PbTiO$_{3}$. For this reason we compare fractional changes in the
structural parameters, rather than absolute values, as n$_s$ and
n$_p$ vary.

\bibitem{Vanderbilt1993}
R. D. King-Smith and D. Vanderbilt, {\sl Phys. Rev. B} \textbf{47}
R1651 (1993).

\bibitem{Tinte2003}
S. Tinte, K.M. Rabe and D. Vanderbilt {\sl Phys. Rev. B.}
\textbf{68} 144105 (2003).

\end{thebibliography}
\end{document}